# Output and citation impact of interdisciplinary networks: Experiences from a dedicated funding program

Nadine Rons

*Nadine.Rons@vub.ac.be*
Vrije Universiteit Brussel (VUB), B-1050 Brussels (Belgium)

**Introduction**

In a context of ever more specialized scientists, interdisciplinarity receives increasing attention as innovating ideas are often situated where the disciplines meet. In many countries science policy makers installed dedicated funding programs and policies. This induces a need for specific tools for their support. There is however not yet a generally accepted quantitative method or set of criteria to recognize and evaluate interdisciplinary research outputs (Tracking and evaluating interdisciplinary research: metrics and maps, 12th ISSI Conference, 2009). Interdisciplinarity also takes on very different forms, as distinguished in overviews from the first codifications (Klein, 1990) to the latest reference work (Frodeman et al., 2010). In the specific context of research measurement and evaluation, interdisciplinarity was discussed e.g. by Rinia (2007) and Porter et al. (2006). This empirical study aims to contribute to the understanding and the measuring of interdisciplinary research at the micro level, in the form of new synergies between disciplines. Investigation of a specialized funding program shows how a new interdisciplinary synergy and its citation impact are visible in co-publications and co-citations, and that these are important parameters for assessment. The results also demonstrate the effect of funding, which is clearly present after about three years.

**Method and material**

The 'Horizontal Research Actions' program was set up at the Vrije Universiteit Brussel in 2002. It supports collaborations joining expertise from different disciplines around topics proposed by the applicants. On average, an application involves four applicants from three departments. Funding is spent primarily on researchers embodying the link between the disciplines. The program was evaluated when the first four generations of 36 applications (incl. 3 resubmissions) could be followed for three years after start of funding. The evaluation used an author-centered approach, based on the applicants and their affiliated departments. Co-publications, defined as joint publications by applicants from different departments, were monitored as an indicator of interdisciplinary output. Co-citations, defined as publications citing applicants from different departments, were monitored as an indicator for citation impact. The basis for analysis was the on line Web of Science. Applications completely situated in the Social Sciences and Humanities (4 out of 36) were excluded due to the insufficient coverage for such networks and remain out of scope of the discussion that follows.

**Results and conclusions**

The results provide information on the program's success as well as on potential indicators for evaluation of interdisciplinary research. The majority of funded applications (9/12) successfully generated both co-publications and co-citations, while about half of the unfunded applications (8/17) showed neither. Despite not being funded, about one third of the unfunded applications (6/17) did also lead to co-publications and co-citations. In the subset of newly activated networks, i.e. where co-publications were not yet present before application, the effect of funding is visible in more strongly rising co-publications and co-citations (Figure 1; 73% citing co-publications; 13% itself co-publications) from the third year.

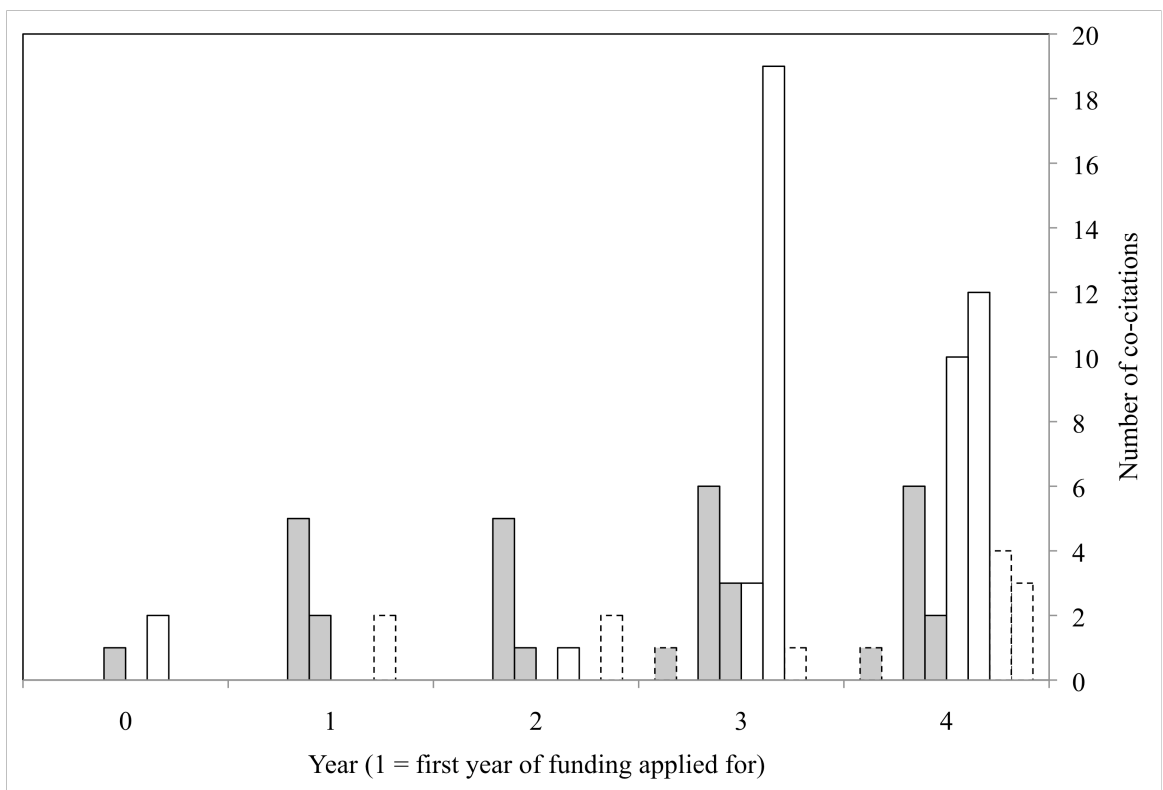

**Figure 1. Co-citations to partners of newly activated interdisciplinary networks**
networks funded for 4 years (white) and unfunded (gray), from corresponding generations
dashed border = with partner in social sciences and humanities

A survey of the five earliest funded networks, all newly activated, confirmed that in line with the program's goal, the large majority of the co-publications represent a synergy of expertise related to the topic (23/28). The remaining concerned rather an application of results from one discipline in another (2/28) or were not related to the topic (3/28). Overall also the majority of the co-citations monitored was related to the topic (83/130), with considerable differences between networks. The non-related co-citations indicate that new collaborations funded by the program may in addition lead to new interdisciplinary combinations of knowledge on another topic, in or outside of the initial network. The survey also showed that the monitored co-publications and co-citations contain the majority of the interdisciplinary output and impact generated by the networks in relation to the topics. This indicates that co-publications and co-citations are important parameters for the assessment of interdisciplinary synergies, e.g. in intermediate evaluations for funding programs after three or more years.